\documentclass[pra,showpacs,amssymb,aps,preprint]{revtex4}

\begin{document}

\title{Time correlation functions and transport coefficients of \\
two-dimensional Yukawa liquids}

\author{Z. Donk\'o$^{1,2}$, J. Goree$^3$, P. Hartmann$^{1,2}$, Bin Liu$^3$}

\address{$^1 $Research Institute for Solid State Physics and Optics of
the Hungarian Academy of Sciences,\\ H-1525 Budapest, P.O. Box 49,
Hungary}

\address{$^2 $Physics Department, Boston College, 
Chestnut Hill, MA 02467, USA}

\address{$^3 $Department of Physics and Astronomy,
The University of Iowa, Iowa City, Iowa 52242, USA}

\date{\today}


\begin{abstract}

The existence of coefficients for diffusion, viscosity and thermal
conductivity is examined for two-dimensional (2D) liquids.
Equilibrium molecular dynamics simulations are performed using a
Yukawa potential, and the long-time behavior of autocorrelation
functions is tested. Advances reported here as compared to
previous 2D Yukawa liquid simulations include an assessment of the
thermal conductivity, using a larger system size to allow
meaningful examination of longer times, and development of
improved analysis methods.
We find that the transport coefficient exists for
diffusion at high temperature, and viscosity at low temperature,
but not in the opposite limits. The thermal conductivity
coefficient does not appear to exist at high temperature. Further
advances in computing power could improve these assessments by
allowing even larger system sizes and longer time series.

\end{abstract}


\maketitle

\section{Introduction}

In strongly coupled dusty plasmas \cite{dusty}, consisting of
micron-sized, highly charged particles, the interaction of the
dust particles in many cases can be well approximated by the
Yukawa potential. Other systems for which this type of potential
is also appropriate include charged colloids \cite{coll} and high
energy density matter \cite{sccs}.

The interparticle Yukawa potential energy in these systems,
\begin{equation}
\phi(r) = \frac{Q^2}{4 \pi \varepsilon_0}
  \frac{\exp(- r / \lambda_{\rm D})} {r},
\label{eq:yukawa_potential}
\end{equation}
accounts for the Coulomb repulsion of the particles originating
from their like charges ($Q$) and the screening of the plasma
which surrounds the dust particles. Screening is characterized by
the dimensionless ratio $\kappa = a / \lambda_{\rm D}$, where $a$
is the Wigner-Seitz radius and $\lambda_{\rm D}$ is the screening
length.

 The coupling parameter $\Gamma$ is a measure of interparticle potential
 energy as compared to kinetic energy. 
 Defined as $\Gamma = Q^2 / (4 \pi \varepsilon_0 a
k_B T)$, it varies inversely with temperature $T$. In the
strong-coupling domain $\Gamma
>$ 1, the system behaves like a non-ideal gas, liquid, and then
solid, as $\Gamma$ increases.

Dusty plasmas in nature and in laboratory environments appear in
both three-dimensional (3D) and two-dimensional (2D) settings. A
notable type of 2D systems is a layer of dust particles levitated
in gaseous discharges. During the past decade this latter system
has been investigated both experimentally and by different
theoretical and simulation approaches. These studies at first
mainly concerned the self-organized crystalline state of the
system \cite{crystal}.  Propagation of compressional and shear
waves has been studied and dispersion properties of these waves
have been determined \cite{crwave}. Using perturbation methods the
generation of Mach cones and nonlinear waves has also been
investigated \cite{mach}. Time scales in these 2D dusty plasma
suspensions are characterized by
$\omega_0^{-1}=(Q^{2}/2\pi\varepsilon_{0}ma^{3})^{-1/2}$, where $m$
is the dust particle mass.

These crystals can be melted, to produce an interesting liquid
state, by changing the plasma parameters or by using laser heating
techniques \cite{heating}. The strongly-coupled liquid state,
established this way, has already attracted the interest in a
series of experiments aimed at the observation of
waves \cite{waves}, as well as studies of transport
processes \cite{NG04}. Apart from experiments, combined theoretical
and simulation studies have recently been carried out in order to
uncover thermodynamic and structural properties \cite{H05} as well
as collective excitations \cite{PRL2004} of two-dimensional Yukawa
liquids. Studies of transport properties in such 2D systems are
especially interesting due to the fact that doubts about the
existence of transport coefficients in low dimensional systems have
been raised on theoretical grounds \cite{AW1970,EHL1970}.

Experiments with dusty plasmas are invariably
nonequilibrium problems, with a constant energy input and
frictional energy loss. The energy input can be provided by ion
flow or electric field fluctuations that propagate into the dust
suspension from an outside source, or by external manipulation
using for example laser radiation pressure. A major source of
friction is gas drag experienced by the solid dust particles as
they move through ambient neutral gas. At steady-state, the
temperature is determined by a balance of the energy input and
dissipation.

Molecular dynamics simulations have successfully been used
for the determination of the self-diffusion \cite{OH00}, the shear
viscosity \cite{SM01,SH02,SC02FM03,Ramazanov,dh2008} and the thermal
conductivity \cite{SC02FM03,DH04} coefficients of 3D Yukawa
liquids.

In general, valid transport coefficients are more likely to exist
in 3D liquids than in 2D. Here we will investigate 2D liquids. The
first question to address is not how large a transport coefficient
is, but rather does it exist at all. Assessing the existence of
diffusion, shear viscosity and thermal conductivity coefficients
in 2D Yukawa systems is the aim of this paper.

Assessing the existence of a transport coefficient in 2D Yukawa
liquids based on simulation data has until now sometimes relied on
qualitative judgments and data that covered too short a time span.
Here we will attempt to improve on both of those limitations. We
will use a larger system size, allowing us to calculate
autocorrelation functions that are meaningful over a longer time
span. We will reduce the qualitative aspect of the judgment by
introducing quantitative analysis methods making use of
Student's t-statistics to conclude whether a correlation function
decays faster than $1/t$. We will also implement other
improvements in the fitting that is part of the analysis, first to
avoid errors that were previously caused by including the initial
decay portions of the autocorrelation function(s), and second to
generate error values required for the Student's t-statistics. Our
assessment in the end cannot be viewed as a final word on the
existence of a transport coefficient, but rather as an improved
estimate that could in the future be improved further, especially
with greater computational power to provide better signal-to-noise
ratios and even longer time spans.

In Sec. II of the paper we introduce the autocorrelation functions to be 
studied and present their general properties, while Sec. III briefly 
discusses previous work. In Sections IV and V, respectively, we explain 
the simulation and data analysis methods. The results are presened in
Sec. VI. Section VII gives the conclusions of our studies.

\section{Autocorrelation functions}

For equilibrium systems, without gradients, transport coefficients
are calculated using the Green-Kubo relations. For the three
transport coefficients of interest here, these are given as
follows \cite{HDbook}.

For the diffusion coefficient, $D$,
\begin{eqnarray}
D  =  \frac{1}{N_d} \int_0^\infty C_v(t) {\rm d}t,& ~C_v(t) \equiv
\langle {\bf v}(t)\cdot {\bf v}(0) \rangle. \label{eq:gk1}
\end{eqnarray}
The integrand $C_v(t)$ is the velocity autocorrelation function
(VACF).

For shear viscosity, $\eta$,
\begin{eqnarray}
 \eta= \frac{1}{V k T} \int_0^\infty C_{\eta}(t) {\rm d}t, & ~~C_{\eta}(t)
\equiv \langle P_{i j}(t) P_{i j}(0) \rangle, \label{eq:gk2}
\end{eqnarray}
where the integrand $C_{\eta}(t)$ is the shear stress autocorrelation
function (SACF).

For thermal conductivity $\lambda$,
\begin{eqnarray}
\lambda =
 \frac{1}{V k T^2} \int_0^\infty C_{\lambda}(t) {\rm d}t, &
~~C_{\lambda}(t) \equiv \langle J_{Q i}(t) J_{Q i}(0) \rangle,
\label{eq:gk3}
\end{eqnarray}
where the integrand $C_{\lambda}(t)$ is the energy current
autocorrelation function (EACF).

Here, $N_d$ is the dimensionality of the system, $V$ is the system
volume, $i \neq j$ are space coordinates, and the brackets
$\langle \rangle$ denote an ensemble average. Calculating the
correlation functions requires time series data for the particle
velocity ${\bf v}$, the off-diagonal element of the pressure
tensor $P_{i j}$, and the energy current $J_{Q i}$; these time
series can all be recorded during a molecular dynamics simulation.

For transport coefficients to exist, the autocorrelation functions
in the integrands of Eqs.~(\ref{eq:gk1})-(\ref{eq:gk3}) must decay
rapidly enough for the integral to converge. This rapid decay for
convergence is the essential requirement that we test in this
paper. In examining their decay, it is crucial to examine the {\it
long-time} behavior. Since the Green-Kubo integrals in
Eqs.~(\ref{eq:gk1})-(\ref{eq:gk3}) extend in principle to
infinity, what is really needed is information about the
integrands as $t \rightarrow \infty$. However, a numerical
simulation is of course limited to a finite time span. This is one
reason that a conclusion regarding the existence of a transport
coefficient cannot be definitive when based on a simulation.
Advances in computational power over the years will allow longer
time spans, so that conclusions regarding the existence of a
transport coefficient can change. Here we take a step in this
gradual improvement by using a bigger simulation size, which
allows a longer time span.

For the purpose of illustrating the autocorrelation functions, we
present some examples in Fig.~\ref{fig:threecurves}. Curves are
shown for $C_v$, $C_{\lambda}$, and $C_{\eta}$, all computed for a
cool liquid ($\Gamma = 300$ and $\kappa = 2$). (Fig.~\ref{fig:threecurves} 
actually shows $|C_v|$, as this function may acquire negative values 
even at ``early'' times due to caged motion of particles.) We plot the 
data on log-log axes so that a power law behavior appears as 
a straight line.

Examining $C_v$ in Fig.~\ref{fig:threecurves}, we identify three
portions of the time series. Our analysis will include only the
central one of these three portions.

First, there are {\it initial decaying oscillations} in the
correlation function. These are associated with caged particle
motion, and are not of interest for diffusion.

In the middle, there is a portion we term {\it smooth decay}. This
portion of the correlation function is what we seek to analyze, by
examining and fitting it to determine whether it decays faster
than $1/t$.

Finally, the smooth decay ends one of two ways: either a {\it
sound peak} as seen in Fig.~\ref{fig:threecurves} for $C_v$ and
$C_{\lambda}$, or {\it noise oscillations with a zero crossing} as
seen here for $C_{\eta}$. The sound peak occurs at a time that is
the ratio of the length  of the simulation box divided by the
sound speed. Correlation data after the sound peak can be
meaningless, especially in a solid or a cold liquid, because of
the periodic boundary conditions, which mean that a sound wave
that exits the box will re-enter the box from the other side. At
long times, when the correlation function has diminished to a
small value, it is obscured by noise, which can appear as
oscillations crossing zero. We will not analyze data after a sound
peak, or a zero crossing, whichever comes first.

\section{Previous work}

In the study of liquids, long-time tails of autocorrelation
functions have been the focus of numerous investigations. These
have included both 3D and 2D systems, with various interparticle
potentials.

Non-exponential long-time tails in the VACF of {\it hard
sphere and hard disk} systems were first reported by Alder and
Wainwright \cite{AW1970}. For the 2D case they observed a
$\propto t^{-1}$ decay of the tail of the VACF, which makes
the VACF non-integrable. As a consequence, the diffusion
coefficient was claimed not to exist for this system.

Further, Ernst {\it et al.} \cite{EHL1970} have shown that the {\it
kinetic contributions} to the autocorrelation functions of shear
stress and energy current -- which are related to velocity
correlations -- exhibit the same behavior. Their findings were
also confirmed by the calculations of Dorfman and
Cohen \cite{DC1972}. Regarding systems with continuous potentials,
power law decay of the VACF was observed in 3D soft-repulsive and
Lennard-Jones liquids \cite{DRS01}. A $t^{-1}$ tail of the SACF was
found in molecular dynamics simulations in the case of 2D soft
disk fluid \cite{ME1985}. For the case of Coulomb interaction (2D
classical electron liquid) the existence of self-diffusion
coeffiicent has been a topic of
controversy \cite{Hansen1979,Baus1980,Agarwal1983}.

For thermal conductivity, the present authors are aware of
experiments  but no previous simulation or theory for a strictly
2D (monolayer) Yukawa liquid. A recent
experimental measurement of the thermal conductivity was reported
for a 2D liquid in \cite{nosenko}, following earlier measurements in
2D crystals \cite{heatexp} and liquids in a quasi-2D
system \cite{Fortov}.

The situation for self-diffusion and shear viscosity in 2D Yukawa
liquids is different, as the literature includes both experimental
and simulation studies. These have yielded estimates of the
self-diffusion coefficient \cite{diffexpt,LG2007} and the shear
viscosity \cite{NG04,G05,Vaulina1,LG05}. Thus, our goals here will
include reporting, apparently for the first time, a simulation for
thermal conductivity in 2D Yukawa liquids, as well as an
improvement over previous simulations for diffusion and viscosity.

For diffusion, earlier equilibrium MD simulations of 2D Yukawa
liquids, restricted to $\kappa$ = 0.56, predicted that
superdiffusion rather than diffusion occurs over a wide range of
temperatures \cite{LG2007}. They also predicted that $D$ exists in
the nonideal gas phase and near the disordering transition
\cite{LGV2006}, and that the Stokes-Einstein relation $D \eta
\propto kT$ is violated very near the disordering transition.

Recent simulations \cite{LG05} motivated by an experiment on the
shear viscosity in 2D Yukawa liquids using a dusty plasma
monolayer \cite{NG04} indicated a fast decay of the SACF that
allowed a calculation of a viscosity coefficient. Non-equilibrium
simulations of the same system \cite{DGHK2006} did not show
significant sensitivity of the results on system size and allowed an
identification of non-Newtonian behavior under high shear rates.

Interest in systems characterized by soft potentials motivates our
investigations of 2D Yukawa liquids. Our aim is to carry out a
systematic study of the time correlation functions and to
investigate the existence of the related transport coefficients of
2D Yukawa liquids, covering a wide range of the $\Gamma$ and
$\kappa$ parameters. Such investigations are especially timely
now as very recent large-scale simulations of Isobe of a 2D hard disk 
fluid systems \cite{Isobe} have demonstrated that for some
conditions the VACF decays {\it slightly but definitely} faster than $1/t$, 
in contradiction with the early findings of Alder and Wainwright 
\cite{AW1970} mentioned above.

Besides analyzing the correlation functions we also check the limits of
applicability of the simulations, which has been missed in some
previous studies. 

\section{Simulation technique}

\subsection{Molecular dynamics method}

It is noted that while nonequilibrium methods are generally more
efficient in studies of transport phenomena, equilibrium simulations
have the advantage of allowing a direct test for the existence of
transport coefficients, as explained above. In cases when this test
indicates that a valid transport coefficient does not exist,
nonequilibrium simulations yield ``transport coefficients'' that may
not be unique, in a way that they exhibit dependence on system size.

The system studied here is strictly two-dimensional, and it is
under equilibrium conditions unlike some experiments, which are
driven-dissipative \cite{NG04}. Using this assumption we integrate
the Newtonian equation of motion of the particles during the
course of the simulation. We apply a rectangular cell with edge
lengths $L_x \cong L_y$, and periodic boundary conditions.
Pairwise Yukawa interparticle forces are summed over a
$\kappa$-dependent cutoff radius (also extending into the images
of the primary computational cell), using the chaining mesh
technique. No thermostat is used, the desired system temperature
is set by rescaling the momenta of the particles in the
initialization phase of the simulation that precedes the start of
data collection.

\subsection{Calculation of functions}

The main results of the simulations are the correlation functions
$C_v(t)$, $C_{\eta}(t)$, and $C_{\lambda}(t)$, defined by
Eqs.~(\ref{eq:gk1})-(\ref{eq:gk3}). In addition to these correlation
functions, to detect diffusion we also calculate time series for the
mean squared displacement of particles
\begin{equation}
{\rm MSD}(t) = \langle ~|{\bf r}(t) - {\bf r}(0) |^2 ~\rangle.
\label{eq:msd}
\end{equation}
The MSD has an advantage, as compared to the VACF, at high
$\Gamma$ where oscillations in the VACF obscure its decay.
Diffusive motion is characterized by a constant time derivative of
the MSD.

Phase space coordinates of the particles allow  the determination of
the VACF and MSD directly, while the time series needed to calculate
the SACF and EACF are obtained from the phase space coordinates as 
\cite{HDbook}:
\begin{eqnarray}
P_{xy} = \sum_{i=1}^N \biggl[ m v_{ix} v_{iy} - \frac{1}{2} \sum_{j
\neq i}^N \frac{x_{ij}y_{ij}}{r_{ij}} \frac{\partial
\phi(r_{ij})}{\partial r_{ij}} \biggr], \label{eq:pxy}\\
J_{Qx} = \sum_{i=1}^N v_{ix} \biggl[ \frac{1}{2} m  |{\bf v}_i|^2 +
\frac{1}{2} \sum_{j \neq i}^N \phi(r_{ij})  \biggr] \nonumber \\
- \frac{1}{2} \sum_{i=1}^N \sum_{j \neq i}^N ({\bf r}_i \cdot {\bf
v}_i) \frac{\partial \phi(r_{ij})}{\partial r_{ij}}, \label{eq:jx}
\end{eqnarray}
where ${\bf r}_{ij} = {\bf r}_i - {\bf r}_j = (x_{ij},y_{ij})$. We
normalize distances by the 2D Wigner-Seitz radius $a=(1/n
\pi)^{1/2}$, where $n$ is the areal density.

The first term on the right hand side of Eq.(\ref{eq:pxy}) is
called the kinetic term, while the second term is the potential
term, i.e.
\begin{equation}
P_{xy} = P_{xy}^{\rm kin} + P_{xy}^{\rm pot}.
\label{eq:pxyterms}
\end{equation}
Similarly, the energy current may be written as
\begin{equation}
J_{Qx} = J_{Qx}^{\rm kin} + J_{Qx}^{\rm pot} + J_{Qx}^{\rm coll},
\label{eq:jqxterms}
\end{equation}
where the three (kinetic, potential, and collision \cite{Eapen} or
virial \cite{AS}) terms correspond to the ones on the right hand
side of Eq.(\ref{eq:jx}). (It is noted that some authors follow a
different partitioning of the energy current.) The stress
autocorrelation function (SACF) thus has the form
\begin{eqnarray}
C_\eta = \langle P_{xy}^{\rm kin}(t)  P_{xy}^{\rm kin}(0) \rangle +
 \langle P_{xy}^{\rm pot}(t)  P_{xy}^{\rm pot}(0) \rangle + \nonumber \\
 2 \langle P_{xy}^{\rm kin}(t)  P_{xy}^{\rm pot}(0) \rangle =
 C_{\eta}^{\rm KK} + C_{\eta}^{\rm PP} + 2 C_{\eta}^{\rm KP}.
\label{eq:cetaterms}
\end{eqnarray}
The energy current autocorrelation function (EACF) may as well
be decomposed as
\begin{eqnarray}
C_{\lambda} = \langle J_{Qx}^{\rm kin}(t)  J_{Qx}^{\rm kin}(0) \rangle +
 \langle J_{Qx}^{\rm pot}(t)  J_{Qx}^{\rm pot}(0) \rangle + \nonumber \\
 \langle J_{Qx}^{\rm coll}(t)  J_{Qx}^{\rm coll}(0) \rangle +
 2 \langle J_{Qx}^{\rm kin}(t)  J_{Qx}^{\rm pot}(0) \rangle + \nonumber \\
 2 \langle J_{Qx}^{\rm kin}(t)  J_{Qx}^{\rm coll}(0) \rangle +
 2 \langle J_{Qx}^{\rm pot}(t)  J_{Qx}^{\rm coll}(0) \rangle = \nonumber \\
 C_{\lambda}^{\rm KK} + C_{\lambda}^{\rm PP} + C_{\lambda}^{\rm CC}+
 2 C_{\lambda}^{\rm KP} + 2 C_{\lambda}^{\rm KC} + 2 C_{\lambda}^{\rm PC}.
\label{eq:clambdaterms}
\end{eqnarray}

In calculating these correlation functions from simulation data,
we use a common method of overlapping time segments. Each
overlapping time segment begins at a different initial time in the
time series. We average the correlation functions over all the
overlapping segments. This averaging serves the role of an
ensemble average. It also serves to reduce the noise.

For the VACF, we are also able to average over all the particles
in the system, and this greatly reduces the noise as compared to
the SACF and EACF, which allow only averaging over the overlapping
segments. For this reason, noise presents a greater challenge for
the SACF and EACF than for the VACF.

We repeat all the simulations several times (the actual numbers will
be given for the cases presented later), each with different
initial conditions for the particle positions. We combine the
results of these runs, yielding a mean value and an error bar for
each data point in the time series for the autocorrelation
function.

\section{Data analysis method}

Here we define in detail our analysis methods for the long-time
behavior of the three autocorrelation functions. Our goal is to
assess whether the functions decay faster than $1/t$, so that
their integrals will converge and the corresponding transport
coefficient exists. This assessment has in the past relied to a
great extent on qualitative inspections of the autocorrelation
functions. The analysis methods described below include
improvements to invoke more quantative criteria for this
assessment. In the end we will answer the question of whether the
transport coefficient exists along with a measure of our
confidence in the conclusion.

\subsection{Fitting the correlation function}

\label{sec:fit}

First, we choose the time range for our analysis. This is done by
inspecting a plot of the correlation function $C(t)$ on log-log
scales, as in Fig.~\ref{fig:threecurves}. We select a starting
time and a stopping time.

The starting time will be selected usually as a judgment of the
time when the initial decay ends. This judgment is qualitative,
which is a concern because we are attempting to reduce the role of
qualitative judgments. Therefore, we will bracket this time and
produce our final results for two or three different starting
times, to judge whether the slightly arbitrary choice of a
starting time has much impact. In some cases, instead of a
qualitative judgment of the initial decay, we will use  a
crossover of the separate terms contributing to the SACF or EACF
to determine the starting time. We believe that in some previous
simulations \cite{LG05}, a starting time was chosen too soon, so
that the analysis included a great deal of the initial decay, so
that the results should not be relied upon. It is only because we
now use a larger simulation size, so that the sound peak occurs
later, that we can detect that this problem occurred in previous
simulations of a smaller size.

The stopping time will be the sound peak or a zero crossing,
whichever comes first. In the example of Fig.~\ref{fig:threecurves},
the analysis will stop at the sound peak for the VACF and EACF, 
but at a zero crossing for the SACF. To detect how sensitive our 
result is to the stopping time, we repeat all our calculations using
$95\%$ of the zero crossing as stopping time.

Second, we fit the data between the starting and stopping times to
a power-law decay,
\begin{equation}
\log [C(t)] = \alpha \log (t) + {\rm{intercept}}.
\label{eq:whatwefit}
\end{equation}
To avoid introducing a bias by overemphasizing data at long times,
before fitting we resample  $C(t)$ at times that are uniformly
spaced when plotted with a logarithmic axis.

Our fit to Eq.~(\ref{eq:whatwefit}) uses a calculation of a
$\chi^2$ surface. The surface is calculated in the parameter space
of the slope $\alpha$ and the intercept. This method requires
error bars for each data point in $C(t)$. We produce these error
bars by performing multiple simulations for the same conditions
but different initial conditions, and calculating the mean and
standard deviation of the mean for each data point in $C(t)$.

An example of a $\chi^2$ surface is shown in
Fig.~\ref{fig:chisquaresurf}. The best fit is the minimum of the
$\chi^2$ surface, yielding $\alpha_{\rm fit}$. The quality of the fit
is evaluated by comparing the minimum $\chi^2$ to $1.00$, which is
considered a typical value for a moderately good fit. In order to
perform further statistical analysis we require an uncertainty, or
standard error, for the fit parameter $\alpha$. We estimate this
uncertainty $\sigma_{\alpha}$ using the $68.3\%$ confidence
interval in the $\chi^2$ surface, as shown in
Fig.~\ref{fig:chisquaresurf}.

\subsection{Hypothesis testing}

To complete our analysis, we use Student's t-statistics to compare
 $\alpha_{\rm fit}$ to $-1$.  First, we compose a
null hypothesis
\begin{equation}
{\rm H0}: \alpha_{\rm fit} > -1 \label{nullhypothesis}.
\end{equation}
In other words, our null hypothesis ${\rm H0}$ is that the
transport is anomalous. We calculate the $t$ value as
\begin{equation}
t = \frac{\alpha_{\rm fit} + 1}{\sigma_{\alpha}}.\label{tvalue}
\end{equation}
Next, we calculate the $p-$value for a one-tailed test using a
Student's t-calculator. The $p-$value is the probability of
$\alpha > -1$. Finally, we determine a significance level, $1-p$,
to reject ${\rm H0}$. Thus, in the end we determine the
significance level for a conclusion that the transport is
anomalous.

To explain this in physical terms, we are attempting to reject the
null hypothesis that the transport coefficient does not exist,
i.e., to reject the idea that the transport is anomalous. If we
find that the significance level for this test is very high, for
example $99\%$, we would be  confident in saying that the
transport coefficient exists. However, if the significance level
is much lower, for example $60\%$, we would be unable to conclude
whether the transport coefficient exists. As always with
Student's t-statistics, the test of the null hypothesis works only
one way. We contemplate whether we can reject ${\rm H0}$, not the
converse. Thus, if we find a small $p-$value such as 0.05 we will
conclude that the transport is not anomalous, with a significance
leve $1-p$. If we instead find a large $p-$value such as 0.40,
the test does not yield a quantitative conclusion but we can
examine the correlation function graph qualitatively for
indications that suggest the transport is anomalous.

\section{Results and discussion}

Most of the correlation function data presented below are based on
several independent simulation runs, each with 1.1 $\times 10^6$ time steps 
(typically spanning a time several times $10^4 \omega_0^{-1}$), 
carried out on systems of $N$ = 4080 or 16~320
particles. After preparing the correlation functions, when we use the
$\chi^2$ surface method, we limit the analysis to two cases. These
are a cool liquid with $\Gamma = 300$ and a warm liquid with
$\Gamma = 20$, both for $\kappa = 2$. These cases correspond to
temperatures of $1.4~T_{M}$ and $21~T_{M}$, respectively, using
data for the melting point $T_M$  from \cite{H05}. Results for the
$\chi^2$ surface and Student's t-method are presented in
Tables~\ref{tab:vactable}-\ref{tab:eacf}.

\subsection{VACF and Diffusion}

To diagnose diffusive motion, we present the analysis of the
velocity autocorrelation functions (VACF) in 
Figs.~\ref{fig:vac}--\ref{fig:sound} for a range of $\Gamma$ and
$\kappa$ values. 
 
\subsubsection{VACF}

We will investigate the decay of the VACF curves plotted in
Fig.~\ref{fig:vac} for $\kappa = 2$. These data represent the average
of 6 simulation runs (for each $\Gamma$) using $N$ = 4080 particles.
The curves have the three
portions described earlier. In the middle there is a smooth decay,
which will be our primary focus. This is preceded by initial
decaying oscillations associated with caged motion, and it is
followed by a sound peak (marked $S$) and noisy oscillations.
Plotting the data on log-log axes reveals a power law, when the
smooth decay appears to be a straight line. For reference, we draw
a line with a slope of $-1$, corresponding to a $t^{-1}$ decay.
Diffusion, as opposed to anomalous diffusion, would require a
decay faster than $t^{-1}$ so that the integral of the VACF
converges.

Our statistical analysis using the $\chi^2$ method is presented in
Table~\ref{tab:vactable}. Two cases are listed here. For a warm
liquid, $\Gamma = 20$, we find that the fit parameter for the
slope is typically  $\alpha \approx 1.20$, depending slightly on
the starting and stopping times used in selecting the data to fit.
Performing the Student's t-analysis leads to the conclusion that
we can reject the null hypothesis ${\rm H0}$ of anomalous
transport with an $83 - 99\%$ significance level. The fits with
the best $\chi^2_{\rm min}$ have a $99\%$ significance level. Thus,
our simulation indicates that the diffusion coefficient exists for
this warm liquid. On the other hand, for a cold liquid, $\Gamma =
300$, $\alpha$ is very near to one, and we cannot reject the null
hypothesis ${\rm H0}$ of anomalous transport.

To further illustrate this trend of diffusive motion, 
we present in Fig.~\ref{fig:alpha}(a) fitting
results for a range of $\Gamma$. (For ease in performing repeated
calculations, here we found $\alpha$ using a simpler fitting
method with only a single set of starting and stopping times, so
that the results are less precise than in Table~\ref{tab:vactable}
and do not have error bars or significance levels.) Results are
reported here for three values of $\kappa$, which all have
different melting points. To better compare these, we rescaled the
horizontal axis as $T/T_M$ in Fig.~\ref{fig:alpha}(b), using data
for the melting point $T_M$ from \cite{H05}. After performing this
rescaling, we find a nearly universal curve: near the melting
point or disordering transition at $T/T_M =1$ there is anomalous
diffusion, while at higher temperatures the motion is  likely to
be diffusive. A transition between the regimes of diffusive and
anomalous particle transport appears at $T/T_M \approx 5$.

\subsubsection{System size effects}

Next we analyze the possible effects related to the finite size of
the computational box. These effects are illustrated here for the
VACF, but the limitations we find are also applicable to the other
correlation functions, the SACF and the EACF.

The finite size of the simulation box limits usefulness of
simulation data in two different ways. First, particles may
traverse a small simulation box without experiencing a sufficient
number of collisions. This may be a concern only for low $\Gamma$
values. Second, sound waves have a finite transit time $\Delta
t_{\rm s}$ across the box \cite{DRS01} as  was already mentioned
above. Due to the periodic boundary conditions, this can limit
meaningful interpretation of correlation functions to $t < \Delta
t_{\rm s}$. This is a concern especially at high $\Gamma$, where
correlations of caged particle motion \cite{cage}, which appear as
oscillations in the correlation function, persist a long time. For
our purposes we wish to interpret correlation functions after
these oscillations have decayed; at high $\Gamma$ this requires a
long time series and thus a large system size. The peaks marked
$S$ in Fig.~\ref{fig:vac} and are conspicuous indications
of this effect.

Random particle motion can be decomposed into a spectrum of sound
waves over a range of wavenumbers. The sound waves have a
dispersion, where the group velocity depends on wavenumber, but
this dispersion is significant only for large wavenumbers
\cite{PRL2004}. For small wavenumbers the wave has little
dispersion, so that there is a distinctive sound speed.

This sound speed depends on the shielding parameter $\kappa$. For
$\kappa$ = 1, 2, and 3, respectively, the sound speeds for such
waves are $c \cong 0.78~a \omega_0$, $c \cong 0.4~a \omega_0$, and
$c \cong 0.22~a \omega_0$ \cite{PRL2004}. The corresponding
 transit times are expected to be $\omega_0
\Delta t_s = \sqrt{N \pi} / c$ = 290, 560, and 950, for $\kappa$ =
1, 2, and 3, respectively, for a simulation box with edge length
$L = a \sqrt{N \pi}$ and $N$ = 16~320 particles.

We see the effect of $\kappa$ in Fig~\ref{fig:sound}(a), where
VACFs show a sound peak that reduces the useful portion of the
data most extremely for small $\kappa$. (The values of the
coupling parameter $\Gamma$ have been varied among the three
curves to keep constant the effective coupling value \cite{H05}
$\Gamma^\star$ = 85.)

The effect of system size on the sound peak is demonstrated in
Fig~\ref{fig:sound}(b). Note that at the smallest system size ($N$
= 1020 particles) the part of the VACF with power law decay
completely disappears. This shows that a large system size is
essential, if one is to identify and fit the decay portion of the
correlation function. This problem becomes more extreme for small
$\kappa$ values where the sound speed is higher [cf.
Fig~\ref{fig:sound}(a)]. For this reason, we believe that our
results represent an advance over some earlier simulations with
small system sizes, where the latter should no longer be relied
upon.

Improving the transit time comes with a significant computational
cost. As $\Delta t_s$ scales as $N^{1/2}$, doubling the useful
long-time range of the VACF requires quadrupling the runtime,
because the number of computations in the simulation scales
linearly with $N$. 

\subsubsection{MSD}

As an additional test for diffusion, to gain confidence in our
conclusions based on the VACF, we examine the long-time behavior
of the mean-square displacement (MSD). This method makes use of
the same kind of simulations, but we use only particle position
data to calculate a time series of squared displacements from an
initial position, and average this over all particles and
overlapping time segments. To do this, we performed additional
simulations with a larger system size of $N$ = 16~320 particles,
but only with a single simulation run. The MSD results are plotted
in Fig.~{\ref{fig:msd}}. First we emphasize that due to the sound
speed, the useful long-time range of the MSD curves is the same as
that for the VACFs. Around the time of $\Delta t_{\rm s}$, marked
$S$, ripples show up on the MSD curves (with an amplitude that is
detectable, but too small to see in Fig.~{\ref{fig:msd}} without
more magnification. These ripples hinder the determination of the
slope at longer times. Thus any analysis of the MSD curves must be
limited to times not exceeding $\Delta t_{\rm s}$.

The MSD time series plotted in Fig.~{\ref{fig:msd}}(a) indicate a
ballistic motion at low $\omega_0 t$ values, characterized by MSD 
$\propto t^2$. At later times the slope of the MSD curves
decays, and for $\omega_0 t \gtrsim 100$ most of the curves appear
to be nearly linear, $\langle (\Delta r)^2 \rangle \propto
t^\gamma$ with slopes $\gamma \gtrsim$ 1.

As a more sensitive indicator, we examine the derivative
$\rm{MSD}'$ of the MSD in Fig.~{\ref{fig:msd}}(b). This derivative
would have a zero slope, at long time, if motion is diffusive.
Instead, we find {\it superdiffusion}, as indicated by an exponent
$\gamma \approx$ 1.15, over the range of $\Gamma \geq $ 100 where
the $\rm{MSD}'$ curves have a constant slope. At $\Gamma < 100$
the slopes of the $\rm{MSD}'$ curves change slightly but
continuously in the time domain extending to $\Delta t_{\rm s}$
[indicated by dotted vertical lines in Fig.~{\ref{fig:msd}}(a) and
(b)]. We speculate that the decreasing slope of the $\rm{MSD}'$ at
lower $\Gamma$ values may eventually, beyond the meaningful time
range shown here, converge to zero, which is required for normal
diffusion. We can, however, not prove this speculation, as several
orders of magnitude longer time may be needed to reach this
convergence. A more convincing demonstration of convergence of the
MSD might require as much as three orders of magnitude additional
$\Delta t_{\rm s}$, which would require $\sim 10^{10}$ particles,
far beyond our computing capacity. Nevertheless, the observation
of a decreasing slope $\rm{MSD}'$ curves at lower $\Gamma$ values
is generally consistent with the findings that the VACFs decay
faster than $t^{-1}$ at these (lower) $\Gamma$ values [cf
Fig.~\ref{fig:vac}(a)], also indicating normal diffusion.

\subsection{SACF and Shear viscosity}

Turning now our attention to the shear viscosity, we present the
stress autocorrelation functions in Fig.~\ref{fig:sacf}. Two cases are shown:
a warm liquid at $\Gamma$ = 20, and a cool liquid at 300, both at $\kappa$ = 2. 
These data are for a large system size of $N = 16~320$, with 108 and 115 
independent simulation runs for $\Gamma = 20$ and 300, respectively,
each comprising 1.1$\times 10^6$ time steps. It is noted that in
the first part of calculations 52 and 31 runs have been carried out
for $\Gamma = 20$ and 300, respectively. The data obtained from these
runs have been analyzed in the way explained in Sec.~\ref{sec:fit},
and the results of this analysis are given in Table~\ref{tab:sactable}.
Having selected the most appropriate start and stop times, several
additional independent simulation runs have subsequently been
carried out (resulting a total number of runs of 108 and 115, for 
$\Gamma = 20$ and 300, respectively) and the data analysis was accomplished 
only for this pair of start and stop times. The best fits obatined
here are also given in Table~\ref{tab:sactable}. (The same procedure was followed 
for the EACF, to be discussed in the next subsection.)

For the warm liquid, $\Gamma = 20$, we observe a power-law tail
with a decay slower than $1/t$. Fitting the decay using our
$\chi^2$ method yields small exponents in the range $-0.63 >
\alpha > -0.80$, as listed in Table~\ref{tab:sactable}. A
Student's t-test yields $p-$values that do not allow rejecting the
null hypothesis of anomalous transport. Examining the SACF graph
in Fig.~\ref{fig:sacf}, we verify that the decay does appear to be
slower than $1/t$. Thus, our simulation indicates that the
viscosity coefficient does not exist for the warm liquid.

This result, that the shear viscosity does not exist for a 2D
Yukawa liquid at a warm temperature, is contrary to what was
previously believed. The difference in our result is presumably
attributable to the larger simulation size, allowing us to observe
the smooth decay after the initial decay.

For the cool liquid, with the higher coupling value of $\Gamma =
300$, we find the opposite result. A qualitative inspection of the
total SACF in Fig.~\ref{fig:sacf} shows a rapid decay, faster than
a power law.

We can gain greater confidence in this conclusion for the cool
liquid by examining the separate contributions to the SACF, in
Fig.~\ref{fig:sacfterms}. These contributions are
 the potential (PP), kinetic (KK) and cross (KP) terms
[as given by Eq.~(\ref{eq:cetaterms})]. Curves shown in
Fig.~\ref{fig:sacfterms} are again for $\Gamma = 20$ and $\Gamma =
300$, at $\kappa = 2$. The values on the vertical axis have been 
normalized so that $C_\eta (t=0)$, 
given as Eq.(\ref{eq:cetaterms}), equals 1.

For the cool liquid at $\Gamma = 300$, the potential term
dominates at early times. More importantly, at longer times we
 see a useful indication that possibly has not been previously
reported: the potential term begins to oscillate, so that it
possibly does not dominate the long-time behavior. At these long
times, $\omega_0 t
> 80$, the kinetic term might instead dominate, and it decays as a power
law. This observation leads us to fit only the kinetic term, and
we find an exponent mostly in the range $-1.29 > \alpha
> -1.38$, and we can reject the null hypothesis of
anomalous viscosity with a significance level of $72 - 99\%$, as
listed in Table~\ref{tab:sacf-kinetic}.

For the warm liquid at $\Gamma = 20$, the kinetic contribution to
the shear viscosity dominates in Fig.~\ref{fig:sacfterms}, as is
well known \cite{SH02,LG05}. In Fig.~\ref{fig:sacfterms} we see
that the potential term decays rapidly into the noise, so that
only the kinetic term contributes significantly to the long-time
tail. Because we see no crossover between the various terms at
long time, we have greater confidence in our conclusion, based on
the total SACF, that the SACF for the warm liquid decays slowly
and the viscosity coefficient does not exist.

\subsection{EACF and Thermal conduction}

Finally we analyze the behavior of the energy autocorrelation
functions (EACF). The EACF for $\kappa$ = 2 at $\Gamma$ = 20 and
300 are plotted in Fig.~\ref{fig:eacf}. These curves are the
results of averaging the same numbers of runs as specified for
the case of the SACF, using $N = 16~320$ particles.

The initial decay persists for a long time in the EACF, as
compared to the  VACF [Fig.~\ref{fig:vac}(a)] or the SACF at low
$\Gamma$ [Fig.~\ref{fig:sacf}(a)]. We judge the initial decay for
the EACF in Fig.~\ref{fig:eacf} to last until the a power law
becomes apparent at $\omega_0 t \approx$ 70 in the case of $\Gamma
= 20$, and $\omega_0 t \approx$ 170 in the case of $\Gamma = 300$.
We will focus our attention now on the ``smooth decay'' that
occurs after this initial decay, and before the first zero
crossing.

For the warm liquid, $\Gamma = 20$, we see in Fig.~\ref{fig:eacf}
a slow power-law decay. As listed in Table~\ref{tab:eacf}, we find
a power-law exponent in the range $-1.16 > \alpha > -1.02$, which
is close to a $t^{-1}$ decay. The $p-$values do not allow
rejecting the null hypothesis of anomalous transport. Examining
the curves qualitatively in Fig.~\ref{fig:eacf}, we conclude that
the decay is likely too slow for convergence. Thus, our simulation
indicates that in this warm liquid, the thermal conductivity
coefficient does not exist.

For the cool liquid, $\Gamma = 300$, our results are hindered by a
short and noisy ``smooth decay'' between the long initial decay
and the oscillations with zero crossings. Fitting, we find in
Table~\ref{tab:eacf} an exponent of $\alpha = -1.40 \pm 0.87$. The
fit is not very reliable, as indicated by the wide error bar on
this value, as well as a high value of $\chi^2_{\rm min}$ in
Table~\ref{tab:eacf}. We are thus unable to conclude whether the
thermal conductivity exists for the cool liquid. Improving this
result would require more extensive computations to improve the
signal-to-noise ratio in the EACF at long times.

We now examine the separate contributions to the EACF in
Fig.~\ref{fig:eacfterms}. For $\Gamma = 20$, there is an apparent
crossover of the KK and CC terms. This crossover may coincide with
the end of the rather long initial decay that we judged in the
total EACF of Fig.~\ref{fig:eacf}. For $\Gamma = 300$, on the
other hand, there is no crossover, giving us more confidence in
our analysis above based on the total EACF.

Our results here for the EACF are apparently the first that have
been reported for a 2D Yukawa liquid. We note that the EACF poses
especially challenging computational requirements because the
useful portion of the curve is at large $t$ when the
signal-to-noise ratio is poorest. 
Improving the signal-to-noise ratio for the EACF at 
$\Gamma$ = 300 to a level that would allow a conclusion 
would require at least an order of magnitude
increase in runs beyond the 115 runs we used.
We concluded that our simulation
indicates the thermal conductivity coefficient does not exist for
a warm liquid, but for a cool liquid, our signal-to-noise ratio
did not allow a conclusion.

\section{Summary}

It has long been suggested, for two-dimensional systems in
general, that valid transport coefficients do not exist
\cite{EHL1970}. Here we reported some counter-examples, based on
our calculations covering a wide range of parameters of 2D
frictionless Yukawa liquids. We find that transport coefficients
sometimes exist, depending on the temperature:

\begin{itemize}

\item{The diffusion coefficient exists for warm but not for cool liquids.
For warm liquids, the diffusion coefficient exists at temperatures
higher than about five times the melting temperature. For cool
liquids, however, we found a closely $\propto t^{-1}$ type decay of the
velocity autocorrelation function (VACF), indicating the occurence
of anomalous diffusion and no valid diffusion coefficient.}

\item{The shear viscosity coefficient exists for cool but not warm
liquids. This finding is contrary to previously reported results.
For $\kappa = 2$, a cool liquid at $\Gamma = 300$ exhibits a fast
decay indicating a valid transport coefficient, but a warm liquid
at $\Gamma = 20$ does not. Comparing to the result above, our
simulations suggest that self-diffusion and viscosity do not
couple, because the coefficients do not exist in the same
temperature regimes.}

\item{The thermal conductivity, assessed here for the first time,
does not exist for a warm liquid at $\Gamma = 20$,
where a slow power-law decay in the energy autocorrelation
function (EACF) was observed. For a cool liquid at $\Gamma = 300$,
however, we are unable to come to a conclusion because of the
technical challenges posed by signal-to-noise ratios and a long
initial decay.}

\end{itemize}

Our approach has been to ask the question, does the transport
coefficient exist? Using equilibrium MD simulations, this question
was addressed by computing autocorrelation functions, and examining
their decay at long times. A rapid decay would indicate that the
integral of the autocorrelation function converges, and the corresponding
 transport coefficient exists. For the results reported here, we
 have improved our statistical analysis methods for assessing
 whether the decay is faster than $1/t$, as required for
 convergence.

A limitation of  equilibrium MD methodology is that testing for
convergence requires reliable measurements of correlation
functions at times tending to infinity, while the simulations
yield results over only a finite time. This finite time is limited
by two challenges: signal-to-noise ratios (which are especially
important for the shear viscosity and thermal conductivity) and
sound peaks arising from periodic boundary conditions (especially
important for the diffusion coefficient). Both of these problems
have been improved here by using larger simulation sizes than in
previous reports. For this paper, we consumed several years of CPU
time, using typical personal computers. Nevertheless, 
the simulation data did not allow us to draw definite conclusions
regarding the existence of transport coefficients in cases
when the decay of correlation functions was close to $1/t$.

Future advances in computational power will allow larger
simulation sizes. The increase in size that is needed is
substantial, because of the square-root scaling of the sound peak
time with respect to system size $N$. A $10^4$ fold increase in
computations would be required for a 100-fold increase in
meaningful time.

Because of these limitations that are gradually being offset by
improved computing power, conclusions like those we presented
above (i) should be considered as a snapshot view of a developing
effort to estimate whether transport coefficients exist in 2D
Yukawa liquids and (ii) may even change just like in the case
of 2D hard disk system (\cite{AW1970} vs. \cite{Isobe}). 
Our results are, nevertheless, valuable  because of interest in
2D liquids in general, and also because of increased interest in
transport measurements in 2D dusty plasma experiments.

We can identify other areas where further work would be useful.
Higher temperature liquids, in the non-ideal gas phase, have not
been well explored yet. Systems present in 2D dusty plasma
experiments -- which include friction and particle heating 
\cite{NG04}, \cite{heatexp} -- can be simulated by
MD methods with proper modification of the particles'
equation of motion. Another line of reaserch could be the
identification of the reasons behind the anomalous transport
and its dependence on the dimensionality of the system. Such
an investigaton has been presented in \cite{sdpaper} for
diffusion.

\acknowledgments

This work was supported by the Hungarian Fund for Scientific
Research and the Hungarian Academy of Sciences, OTKA-T-48389,
OTKA-IN-69892, MTA-NSF-102, OTKA-PD-049991. J.G. and B. L. were
supported by NASA and DOE.

\clearpage

\begin{table}
\caption{\label{tab:vactable} Results for fitting the $C_v$, the
velocity autocorrelation function (VACF). Various starting and
stopping times, normalized by $\omega_0$, are tested. For the warm
liquid, $\Gamma = 20$, the $p$-values are small, leading us to
reject ${\rm H0}$ with an $80 - 99\%$ significance level and to
conclude that the diffusion coefficient exists. For the cool
liquid, $\Gamma = 300$, the $p$-values do not allow us to conclude
anything. $\kappa$ = 2. $N_{\rm R}$ is the number of runs 
included in the data analysis.}
\begin{tabular}{ccccccccc}
  $\Gamma$ & Stop & Start & $\alpha_{\rm fit}$
  & $\sigma_{\alpha}$ & $\chi^2_{\rm min}$ & $t$ & $p$ & $N_{\rm R}$\\
\hline
20 &190 &20  &-1.20 &0.05 &0.92 &4.00 &0.000 &6\\ 
20 &190 &30 &-1.22&0.09 &0.20 &2.59 &0.005 &6\\ 
20 &190 &45 &-1.20 &0.15 &0.20 &1.38 &0.08 &6\\ 
20 &190 &68  &-1.17 &0.20 &0.20 &0.85 &0.20 &6\\ 
20 &180 &20  &-1.20 &0.05 &0.94 &4.00 &0.00 &6\\ 
20 &180 &30  &-1.22 &0.09 &0.20 &2.44 &0.01 &6\\ 
20 &180 &45  &-1.20 &0.15 &0.21 &1.38 &0.08 &6\\ 
20 &180 &68  &-1.19 &0.20 &0.21 &0.95 &0.17 &6\\ 
300 &236 &100 &-1.00 &0.16 &2.60 &0.00 &0.50 &6\\ 
300 &236 &120  &-1.05 &0.44 &0.54 &0.11 &0.45 &6\\ 
300 &236 &148  &-1.02 &0.39 &0.39 &0.05 &0.48 &6\\ 
300 &224 &100  &-1.00 &0.16 &2.70 &0.00 &0.50 &6\\ 
300 &224 &120  &-1.04 &0.22 &0.57 &0.19 &0.43 &6\\ 
300 &224 &148  &-1.02 &0.41 &0.41 &0.05 &0.48 &6
\\
\hline
\end{tabular}
\end{table}

\begin{table}
\caption{\label{tab:sactable} Results for fitting the $C_{\eta}$,
the stress autocorrelation function (SACF). Various starting and
stopping times, normalized by $\omega_0$, are tested for the warm
liquid, $\Gamma = 20$. The $p-$values are high, indicating that we
cannot reject ${\rm H0}$. $\kappa$ = 2. $N_{\rm R}$ is the number of 
runs included in the data analysis.}
\begin{tabular}{cccccccccc}
  $\Gamma$ & Stop & Start & $\alpha_{\rm fit}$
  & $\sigma_{\alpha}$ & $\chi^2_{\rm min}$ & $t$ & $p$ & $N_{\rm R}$\\
\hline
20 &120 &30 &-0.71 &0.84 &0.67 &0.35 &0.64 &52\\ 
20 &114 &30  &-0.67 &0.82 &0.69 &0.40 &0.66 &52\\ 
20 &120 &20  &-0.65 &0.59 &0.56 &0.59 &0.72 &52\\ 
20 &114 &20 &-0.63 &0.70 &0.57 &0.53 &0.70 &52\\ 
20 &120 &13 &-0.79 &0.47 &0.75 &0.45 &0.67 &52\\ 
20 &114 &13  &-0.80 &0.52 &0.76 &0.39 &0.65 &52\\
\hline
20 &120 &20  &-0.69 &0.97 &0.21 &0.32 &0.63 &108
\\
\hline
\end{tabular}
\end{table}

\begin{table}
\caption{\label{tab:sacf-kinetic} Results for fitting only the
kinetic contribution to the stress autocorrelation function (SACF)
for the cool liquid, $\Gamma = 300$. The $p-$values are small, so
that we reject ${\rm H0}$. $\kappa$ = 2. $N_{\rm R}$ is the number of runs 
included in the data analysis.}
\begin{tabular}{ccccccccc}
  $\Gamma$ & Stop & Start & $\alpha_{\rm fit}$
  & $\sigma_{\alpha}$ & $\chi^2_{\rm min}$ & $t$ & $p$ & $N_{\rm R}$\\
\hline
300 &300 &20 &-1.38 &0.19 &1.77 &2.00 &0.02 &31\\ 
300 &300 &30 &-1.36 &0.31 &0.91 &1.16 &0.12 &31\\ 
300 &300 &45  &-1.29 &0.49 &0.95 &0.60 &0.27 &31\\ 
300 &285 &20 &-1.38 &0.19 &1.79 &2.05 &0.02 &31\\ 
300 &285 &30 &-1.36 &0.31 &0.92 &1.18 &0.12 &31\\ 
300 &285 &45  &-1.29 &0.49 &0.96 &0.59 &0.28 &31
\\
\hline
300 &300 &45  &-1.60 &0.74 &0.21 &0.81 &0.21 &115\\ 
\hline
\\
\end{tabular}
\end{table}

\begin{table}
\caption{\label{tab:eacf} Results for fitting the energy
current autocorrelation function (EACF). Compared to the VACF and
SACF, for the EACF we used fewer combinations of start and stop
times due to the limited time duration between them. For $\Gamma =
300$, due to noise and a limited useful time range, the fit
quality as indicated by $\chi^2_{\rm min}$ is not good, and we cannot
rely strongly on the corresponding $p-$value. $\kappa$ = 2. 
$N_{\rm R}$ is the number of runs included in the data analysis.}
\begin{tabular}{ccccccccc}
  $\Gamma$ & Stop & Start & $\alpha_{\rm fit}$
  & $\sigma_{\alpha}$ & $\chi^2_{\rm min}$ & $t$ & $p$ & $N_{\rm R}$\\
\hline
20 &400 &68 &-1.12 &0.60 &1.90 &0.20 &0.42 &52\\ 
20 &360 &68 &-1.02 &0.61 &0.69 &0.03 &0.49 &52\\ 
300 &535 &170  &-1.40 &0.87 &3.21 &0.46 &0.32 &31\\
\hline
20 &360 &68 &-1.16 &0.91 &0.22 &0.18 &0.43 &108\\ 
300 &535 &170  &-1.20 &0.47 &1.26 &0.42 &0.34 &115\\
\hline
\end{tabular}
\end{table}

\clearpage

\begin{figure}[h!]
\caption{(Color online) Examples of the three autocorrelation functions, $C(t)$.
These are the VACF ($C_v$), the SACF ($C_{\lambda}$), and the EACF
($C_{\lambda}$). (The curves are shifted vertically for the clarity of
the plot.) The corresponding transport coefficients
(self-diffusion, shear viscosity, and thermal conductivity,
respectively) are deemed to exist, if $\int C(t) {\rm d}t$ converges.
Data here are from our simulation of a 2D Yukawa liquid at a low
temperature, $\Gamma = 300$. Later, we will analyze data after the
initial decaying oscillations but before a sound peak or zero
crossing.} \label{fig:threecurves}
\end{figure}

\begin{figure}[h!]
\caption{(Color online) A $\chi^2$ surface is used in our analysis
method, for fitting a correlation function from the simulation to
Eq.~(\ref{eq:whatwefit}). For a value of the slope $\alpha$ and
intercept, we calculate the sum of square residuals to find
$\chi^2$, and we repeat
 for many pairs of these values. The resulting values are plotted
  as contours. The best fit is the minimum of the $\chi^2$
  surface. A confidence interval
  (-0.30 to -1.32 for the fit parameter $\alpha$ in this example)
  is found using the contour at a height of 2.3 above the minimum.
  Data shown here are for the SACF, with start and stop
  times of $13$ and $114~\omega_0^{-1}$, respectively.
  $\Gamma=20$ and $\kappa=2$.} \label{fig:chisquaresurf}
\end{figure}

\begin{figure}[h!]
\caption{(Color online)
Absolute value of the velocity autocorrelation functions of 2D
Yukawa liquids at $\kappa$ = 2. The heavy straight line indicates $t^{-1}$ decay.
The curves are shifted vertically for the clarity of the plot.
``S'' marks the spurious sound-peak feature  from the traverse of
the sound wave through the finite-size simulation box.}
\label{fig:vac}
\end{figure}

\begin{figure}[h!]
\caption{(Color online) (a) Exponent $\alpha$ characterizing the
decay of the VACFs obtained from fitting straight lines to the
linearly falling parts of the curves seen in
Fig.~\ref{fig:vac}(a). Data for additional values of $\kappa$ are
also shown. (b) $\alpha$ as a function of the normalized
temperature $T' = T / T_{\rm M}$ appears to be a nearly universal
curve, with anomalous diffusion for $T / T_{\rm M}$ below about
$5$.} \label{fig:alpha}
\end{figure}

\begin{figure}[h!]
\caption{(Color online) (a) Effect of $\kappa$ and (b) system size
(particle number $N$) on the velocity autocorrelation functions of
2D Yukawa liquids. In (a) the coupling coefficient values are
$\Gamma$ = 100, 225, and 656, respectively, for $\kappa$ = 1, 2,
and 3. System size effects indicated by the
sound peak $S$ are most severe for small $\kappa$. The curves are 
shifted vertically for the clarity of the plots.}
\label{fig:sound}
\end{figure}

\begin{figure}[h!]
\caption{(Color online) (a) Mean square displacement MSD and (b)
its time derivative for different $\Gamma$ values at $\kappa$ = 2
and $N$ = 16~320. The heavy lines indicate power law behavior. The
dotted vertical lines marked with ``S'' indicate the maximum valid
time set by the sound speed.} \label{fig:msd}
\end{figure}

\begin{figure}[h!]
\caption{(Color online) Stress autocorrelation functions for
$\Gamma$=20 and 300, $\kappa$ = 2. The curves are shifted vertically 
for the clarity of the plot. The heavy line is a power-law fit 
to the $\Gamma$ = 20 curve with start time = 20 $\omega_0 t$
and stop time = 120 $\omega_0 t$. $N$ = 16~320.}
  \label{fig:sacf}
\end{figure}

\begin{figure}[h!]
\caption{(Color online) Contributions of different terms [see
Eq.(\ref{eq:cetaterms})]
  to the stress autocorrelation
  functions for (a) $\Gamma$=20 and (b) 300, $\kappa$ = 2. $N$ = 16~320.
  $C_{\eta}^{\rm KP}$ is multiplied by $-1$ in (a). Start time = 
  45 $\omega_0 t$, stop time = 300 $\omega_0 t$ for the fit to the
  curve at $\Gamma$ = 300.}
  \label{fig:sacfterms}
\end{figure}

\begin{figure}[h!]
\caption{(Color online) Energy autocorrelation functions for
$\Gamma$ = 20 and 300, $\kappa$ = 2, and $N$ = 16~320. The curves are 
shifted vertically for the clarity of the plot. Start time = 
68 $\omega_0 t$, stop time = 360 $\omega_0 t$ for the fit to the
curve at $\Gamma$ = 20.}
\label{fig:eacf}
\end{figure}

\begin{figure}[h!]

\caption{(Color online) Contributions of the different terms [see
Eq.(\ref{eq:clambdaterms})]
  to the energy autocorrelation functions for $\Gamma$ = 20 (a) and 300 (b),
  $\kappa$ = 2, and $N$ = 16~320.}
  \label{fig:eacfterms}
\end{figure}

\clearpage

\end{document}